\documentclass{osa-article}
\usepackage{caption}
\usepackage{subcaption}
\usepackage{grffile}
\usepackage{booktabs}
\usepackage{multirow}
\journal{osajournal}

\setcopyright{Optical Society of America. One print or electronic copy may be made for personal use only. Systematic reproduction and distribution, duplication of any material in this paper for a fee or for commercial purposes, or modifications of the content of this paper are prohibited.}
\newcommand{\boldr}{\ensuremath{\boldsymbol{r}}}

\newcommand{\localLambda}{\ensuremath{\overleftrightarrow{\Lambda}}}
\newcommand{\defectvol}{\ensuremath{\mathcal{V}_{d}}}
\articletype{Research Article}

\begin{document}

\title{Fast Computation of Scattering by Isolated Defects in Periodic Dielectric Media}

\author{Kuljit S. Virk\authormark{1}}

\address{\authormark{1}KLA, One Technology Drive, Milpitas, CA 95035, USA}

\email{\authormark{*}kuljit.virk@kla.com} 

\homepage{https://www.linkedin.com/in/kuljitvirk/} 

\begin{abstract}
Scattering by an isolated defect embedded in a dielectric medium of two dimensional periodicity is of interest in many sub-fields of electrodynamics. Present approaches to compute this scattering rely either on the Born approximation and its quasi-analytic extensions, or on \emph{ab-initio} computation that requires large domain sizes to reduce the effects of boundary conditions. The Born approximation and its extensions are limited in scope, while the ab-initio approach suffers from its high numerical cost. In this paper, I introduce a hybrid scheme in which an effective local electric susceptibility tensor of a defect is estimated by solving an inverse problem efficiently. The estimated tensor is embedded into an S-matrix formula based on the reciprocity theorem. With this embedding, the computation of the S-matrix of the defect requires field solutions only in the unit cell of the background. In practice, this scheme reduces the computational cost by almost two orders of magnitude, while sacrificing little in accuracy. The scheme demonstrates that statistical estimation can capture sufficient information from cheap calculations to compute quantities in the far field. I outline the fundamental theory and algorithms to carry out the computations in high dielectric contrast materials, including metals. I demonstrate the capabilities of this approach with examples from optical inspection of nano-electronic circuitry where the Born approximation fails and the existing methods for its extension are also inapplicable. 
\end{abstract}
\section{Introduction}

Scattering of electromagnetic waves by objects small compared to wavelength is central to the study of light matter interactions. Scattering of light by small particles suspended in free space has also been studied extensively in astrophysics and atmospheric science for well over a century\cite{hulst1981light}, and its study continues to be of prime importance in microscopy, nano-photonics, and non-invasive sensing. A closely related field to microscopy is the optical inspection of defects in semiconductor chip manufacturing, where light scattered by sub-wavelength structures carries a signal for the presence of a defect that may disable the entire chip. This anomaly detection  far below the diffraction limited resolution plays an important role in sustaining the economy of dimensional scaling of semiconductor devices.

The most widely used approximation in scattering calculations in optics is the Born approximation\cite{hulst1981light,born_wolf_bhatia_clemmow_gabor_stokes_taylor_wayman_wilcock_1999,jackson1999classical,kristensson2016scattering}, in which the solution to Maxwell equations is assumed known for a background dielectric profile, $\varepsilon^{(0)}(\boldr)$. In the presence of a scattering object, the dielectric profile is replaced by $\varepsilon(\boldr)$ and the field is expanded in terms of the perturbation, $\varepsilon(\boldr)-\varepsilon^{(0)}(\boldr)$. This \textit{Born series} is a formally exact solution for the total electromagnetic field, while in the Born approximation, only the first order term is retained. This corresponds to the assumption that the field inside the scattering object changes by small amount from the background field in the absence of that object \cite{hulst1981light}. When this phase shift is not small, the approximation is not guaranteed to be accurate. The problems discussed in this paper deal with this regime.

A less well known first order approximation is the \textit{Rytov approximation}\cite{born_wolf_bhatia_clemmow_gabor_stokes_taylor_wayman_wilcock_1999} in which the solution is sought as a first order correction to the complex phase of a scalar field, $U(\boldr)=e^{i(\psi^{(0)}+\psi^{(1)}+\psi^{(2)}+\ldots)}$. The exponentiation of the phase represents a partial re-summation of the Born series, and the resulting expression may hold over a much larger propagation distance in some scenarios. The validity of Rytov approximation has not been established in as general and rigorous terms as for the Born approximation \cite{Keller:69}. 

The Rytov approximation written in the form  $U(\boldr)=e^{i(\psi^{(1)}+\psi^{(2)}+\ldots)}U^{(0)}(\boldr)$  may be viewed as a special case of a more general class of approximations in the form of linear complex transformations of the background field inside the object,
\begin{equation}
E_i(\boldr)=\sum_{j=1}^3\mathcal{L}_{ij}(\boldr)E_j^{(0)}(\boldr),\label{eq:localfield1}
\end{equation}
where the tensor $\mathcal{L}(\boldr)$ is a complex rotation and scaling of the field locally at each point, and the indices $i,j$ are the Cartesian components. These may also be taken as the polarization components depending on the representation adopted for the fields. This \textit{local field correction} in the limit of infinite wavelength for spherical object is the well-known Clausius-Mossotti or the Lorentz-Lorenz theory of molecular polarizability \cite{jackson1999classical}. 

The form \eqref{eq:localfield1} also arises from a partial re-summation of Born series, and was fruitfully explored by Habashy \cite{Habashy1987,Habashy1993} and Torres-verdin \cite{Torres-verdin1994}, and later expanded for source proximity effects by Abubakar and Habashy\cite{Abubakar2005}. The analysis of the Born series in these works led the authors to a quasi-analytic form for the linear transformation $\mathcal{L}$ relating the scattered field to the background. While the analytical form provides insight into the nature of the approximation, its practical use is limited to the cases where the Green function of the background medium is known (see Section \ref{subsec:susceptibility expansion}). As such the method has been useful in geophysics applications of sub-surface detection in otherwise homogeneous earth in the microwave and radio frequency regimes. 

Similar to the Torres-verdin and Habashy's technique, Zhdanov and Fang \cite{Zhdanov2002} introduced a numerical optimization technique that exploits self-consistency of the integral field equations to evaluate the local field correction tensor. This approach also relies on evaluating the Green function of the background medium within the defect region as well as outside of it. Thus their direct use in the optical and photonics applications is prevented by the tremendous cost of computing this Green function in 3-dimensional complex structured media such as photonic crystals and patterned semiconductor device wafers.

When the approximations above are inadequate, one resorts to solving Maxwell equations \textit{ab-initio} using numerical methods such as finite difference time domain (FDTD), coupled mode theory (RCWA), or the variants of finite element method. To use these methods for computing the field scattered by an isolated scatterer in an otherwise periodic medium, the problem is formulated over a computational domain, called the \emph{supercell} consisting of many unit cells. The unit cell at the center of the supercell is modified to contain the scatterer while the remaining cells are left unperturbed. Bloch's theorem is used to remove domain termination effects, and thus the scatterer is repeated periodically with a period equal to the supercell size. As the size of the supercell is increased the interaction between neighboring defects is diminished and the isolated scatterer limit is reached. Due to the $O(N^2)$ scaling with increase in the area of the cell, the supercell computation can be prohibitively slow. In addition, the supercell must also be large enough to properly sample the incidence directions for the S-matrix. Such computations are prohibitive even with reasonable resources (see Sec. \ref{sec:discussion} ).

The scheme introduced in this paper is also inspired by the linear transformation in \eqref{eq:localfield1}, and resembles the method by Torres-verdin and Habashy. Unlike Torres-verdin and Habashy, I do not derive the localized susceptibility tensor directly from the knowledge of the Green function of the background medium. Instead, I recast the problem in which the transformation tensor elements are treated as unknown parameters, and then develop a numerical technique that relies only on small domain computations to estimate these parameters. Thus we break free of \textit{both} the long computation times of large domain ab-initio simulations, \textit{and} the explicit evaluation of the background Green function \cite{Habashy1987,Habashy1993}. In other words, the costly forward computations of Green function for structured media are replaced by an inverse problem that is far more efficient to solve. The accuracy of this technique increases straightforwardly with more numerical data.

For practical applications, the key distinction of this work lies in accurately estimating the matrix element of scattering of plane waves. The incoming waves travel from infinity towards the dielectric medium and the outgoing waves propagate from the structure towards infinity. Therefore, only the \emph{projection} of the S-matrix onto these traveling modes of half spaces is required. The work exploits the fact that in many cases of interest, obtaining such matrix elements requires far less computational effort than a numerically exact solution of the Maxwell equations. It is important to note that the formalism presented is not restricted to this projection alone. However, the numerical examples in this work demonstrate the gain in computational efficiency for this case only.

Furthermore, the method in this paper works on top of the numerical solutions to Maxwell equations. Therefore, methods that speed up the calculations of the near fields will reduce the absolute computational time of this technique by the same amount. This is so because the gain in efficiency is obtained by reducing the number of independent solutions to Maxwell equations needed to populate the relevant sectors of the S-matrix. The computational domain size needed to compute the relevant near fields is also reduced in cases of high absorption inside the structures. As a result, the technique directly benefits from any advancements made in the near field calculations. 

It may be the case that a near field computation in a particular case is so efficient that the absolute time for computation is acceptable. In that case, the method of this work will not be relevant. However, it is highly likely that this method will remain relevant for a wide range of applications. For example, the practical applications in semiconductor device inspections present a wide range metals, dielectrics, and insulators. Dielectric profiles must be modeled to nanometer scale due to shrinking defect sizes, and the computational domain must exceed a few micrometers to compute optical images. While a near field calculation specialized to a particular structure may provide an efficient path, it is unlikely that techniques other than \emph{ab-initio} methods such as FDTD, Finite Element Methods, or RCWA will have a universal applicability. 

Recent work by Vettenburg \textit{et al.} \cite{Vettenburg:19} introduces a method that derives from the same formalism discussed in this paper, but proceeds to an efficient calculation of near fields. This advancement in solving Maxwell equations will clearly benefit the first step of the computational strategy described in this work, and as discussed briefly in the paragraph above. However, the intricate of the structures in nano-scale logic and memory devices require very high order Fourier modes for applying the Green functions in the diagonal form. The dielectric contrasts between insulators and metals (such as W, Co, Mo) in the ultraviolet to visible range wavelengths result in large discontinuities of the electric field at interfaces. As a result, while the preconditioned series may in fact be convergent in principle, it may require an exceedingly large number of iterations to produce an accurate solution. However, I note that moderate dielectric contrasts are indeed found at various stages of memory device fabrication where the convergent reformulation of Born series may be a very attractive alternative. 

The paper is organized into 3 major sections. In Sec. \ref{sec:theory}, I introduce the formalism defining a susceptibility expansion from which numerical approximations schemes are derived. In Sec. \ref{sec:parameter estimation}, I introduce  parameter estimation of the susceptibility tensor. In Sec. \ref{sec:results},  I describe simplified examples from practical applications from  semiconductor device manufacturing where the aforementioned techniques in the literature become untenable. The examples are simplified for clarity of presentation and also to ensure that only public domain knowledge of the wafer processing is discussed. In Sec. \ref{sec:discussion}, I discuss the reduction in computational effort for sampling the scattering matrix of an isolated defect offered by this method, and potential applications in other fields.

\section{Theory}\label{sec:theory}
\subsection{Non-local susceptibility expansion and local approximations}\label{subsec:susceptibility expansion}
Let us begin by defining a periodic dielectric structure without defects as characterized by linear optical susceptibility, $\chi^{({0})}_{ij}(\boldr,\omega)$ that is a function of angular frequency $\omega$ and position $\boldr$. Let us also define an arbitrary Bravais lattice in two dimensions as generated by a set of basis vectors $\boldsymbol{a}_1$ and $\boldsymbol{a}_2$ such that an infinite set of vectors $\boldsymbol{R}=n\boldsymbol{a}_1+m\boldsymbol{a}_2$, where $n,m$ are integers,  generate all equivalent positions in the lattice. The equivalence of the positions is encoded in the condition, 
\begin{equation}
    \chi^{(0)}(\boldsymbol{r}+\boldsymbol{R},\omega)=\chi^{(0)}(\boldsymbol{r},\omega).\label{eq:susceptibility-1}
\end{equation}
In the dimension orthogonal to the lattice (typically denoted $z$), the structure is sandwiched between half spaces of uniform dielectric. In this work, the upper half space will always be taken as vacuum and will contain both the source and the detectors of electromagnetic radiation. The lower half space represents the substrate (typically Silicon) unless transmission is computed, in which case it is also typically vacuum. The susceptibility in \eqref{eq:susceptibility-1} defines a constitutive relation between the electric field and the polarization, 
\begin{equation}
    P^{(0)}_i(\boldsymbol{r},\omega) = \epsilon_0\chi_{ij}^{(0)}(\boldsymbol{r},\omega) E^{(0)}_j(\boldsymbol{r},\omega),\label{eq:local-constitutive}
\end{equation}
where $\epsilon_0$ is the free space permittivity and summation over repeated indices is implied. The electric field $\boldsymbol{E}^{(0)}(\boldsymbol{r},\omega)$ is given by,
\begin{equation}
E^{(0)}_i(\boldsymbol{r},\omega)=\frac{k^2}{\epsilon_0} \int \mathrm{d}^3\boldsymbol{r}' G^{(0)}_{ij}(\boldsymbol{r},\boldsymbol{r}';\omega) P^{s}_{j}(\boldsymbol{r}',\omega),\label{eq:E0-G-P}
\end{equation}
where $-i\omega\boldsymbol{P}^{s}(\boldsymbol{r},\omega)$ is the source current positioned far outside the structure, and is chosen to be in the upper half space without loss of generality. The source term of main interest in the paper is an infinitely thin sheet generating a plane wave (see Appendix \ref{app:sources} for the exact formula). The \textit{background} Green function $G^{(0)}_{ij}$ has the dimensions of inverse length, and satisfies the equation,
\begin{equation}
    \nabla\times\nabla\times
    G^{(0)}_{ij}(\boldsymbol{r},\boldsymbol{r}';\omega)-
    k^{2}\left[\delta_{il}+\chi^{(0)}_{il}(\boldsymbol{r},\omega)\right]
    G^{(0)}_{lj}(\boldsymbol{r},\boldsymbol{r}';\omega) = \delta(\boldsymbol{r}-\boldsymbol{r}')\delta_{ij}. \label{eq:zeroth-GF-waveq}
\end{equation}
In the above equation, $k^2 = \omega^2/c^2$, where $c$ is the speed of light in vacuum. 

A scatterer, which I call a \textit{defect} in this paper, is modeled via a perturbation in the susceptibility as follows,
\begin{equation}
 \chi_{ij}(\boldr,\omega)=\chi^{(0)}_{ij}(\boldr,\omega)+\chi^{(d)}_{ij}(\boldr,\omega),\label{eq:total-chi}
\end{equation} 
where $\chi^{(d)}_{ij}(\boldr,\omega)$ is zero everywhere except inside the volume $V^d$ of the defect. In the presence of the defect, the total self-consistent field satisfies the equation,
\begin{equation}
    \nabla\times\nabla\times
    E_{i}(\boldsymbol{r},\omega)-
    k^{2}\left[\delta_{ij}+\chi^{(0)}_{ij}(\boldsymbol{r},\omega)\right]
    E_{j}(\boldsymbol{r},\omega) = k^2 \chi^{(d)}_{ij}(\boldsymbol{r},\omega)E_j(\boldr,\omega) + 
    \frac{k^2}{\epsilon_0} P^s_{i}(\boldr,\omega). \label{eq:total-efield-eq}
\end{equation}
This equation describes the total field with the same propagator as the background field, but driven by a self-consistent additional polarization. 
As shown in Appendix \ref{app:E-equation}, the formal solution to the total field can be expressed with a different Green function of dimension inverse volume, such that,
\begin{equation}
    E_{i}(\boldsymbol{r}) = 
    \int \mathrm{d}^3\boldsymbol{r}'
    \mathcal{R}^{(d)}_{ij}(\boldsymbol{r},\boldsymbol{r}')
    E^{(0)}_{j}(\boldsymbol{r}'),\label{eq:E-formal-1}
\end{equation}
where I have suppressed the argument $\omega$ for brevity. The solution is stated in terms of the resolvent operator $\mathcal{R}^{(d)}_{ij}(\boldr,\boldr')$ satisfying an integral equation, written in the dyadic notation as,
\begin{equation}
    \overleftrightarrow{\mathcal{R}}^{(d)}(\boldsymbol{r},\boldsymbol{r}') = 
    \overleftrightarrow{I}\delta(\boldsymbol{r}-\boldsymbol{r}')+
    k^2\int_{V^d} \mathrm{d}^3\boldsymbol{r}''
    \overleftrightarrow{G}^{(0)}(\boldsymbol{r},\boldsymbol{r}'')
     \overleftrightarrow{\chi}^{(d)}(\boldsymbol{r}'')
    \overleftrightarrow{\mathcal{R}}^{(d)}(\boldsymbol{r}'',\boldsymbol{r}').\label{eq:resolvent-IE-1}
\end{equation}
Noting the multiplication by the perturbed susceptibility in the second term, let us define a non-local susceptibility \emph{density},
\begin{equation}
    \overleftrightarrow{\Gamma}^{(d)}(\boldsymbol{r},\boldsymbol{r}') = 
    \overleftrightarrow{\chi}^{(d)}(\boldsymbol{r})
    \overleftrightarrow{\mathcal{R}}^{(d)}(\boldsymbol{r},\boldsymbol{r}').
    \label{eq:gamma-R}
\end{equation}
Substituting \eqref{eq:resolvent-IE-1} into \eqref{eq:gamma-R}, the tensor $\Gamma^{(d)}$ satisfies the equation,
\begin{equation}
    \overleftrightarrow{\Gamma}^{(d)}(\boldsymbol{r},\boldsymbol{r}') = 
    \overleftrightarrow{\chi}^{(d)}(\boldsymbol{r})\delta(\boldsymbol{r}-\boldsymbol{r}')+
    k^2
    \overleftrightarrow{\chi}^{(d)}(\boldsymbol{r})
    \int_{V^d} \mathrm{d}^3\boldsymbol{r}''
    \overleftrightarrow{G}^{(0)}(\boldsymbol{r},\boldsymbol{r}'')
    \overleftrightarrow{\Gamma}^{(d)}(\boldsymbol{r}'',\boldsymbol{r}').\label{eq:gamma-IE-1}
\end{equation}

As all quantities defining the tensor are independent of the source term, the tensor itself is an intrinsic property of the medium and remains the same for any source. I exploit this property in the next section to estimate the tensor from a small number of ab-initio calculations of the total field with different excitation sources. 

Furthermore, since by definition $\chi^{(d)}(\boldr)=0$ for $\boldr$ outside the volume of the defect, successive iteration of \eqref{eq:gamma-IE-1} shows that $\Gamma^{(d)}_{ij}(\boldsymbol{r},\boldsymbol{r}')$ is non-zero only when \textit{both} its arguments lie inside the defect volume as well, \textit{i.e.} $\boldr, \boldr' \in $ defect. Thus each term in the series expansion of $\Gamma^{(d)}(\boldr, \boldr')$ represents background propagation between all pairs of points inside the defect, interrupted by zero to infinitely many scattering events by the defect perturbation. The dependence of the non-local susceptibility on the scattering of the fields by the medium outside is contained fully in the zeroth-order Green function $G_{ij}^{(0)}(\boldr,\boldr')$, the exact computation of which includes all the possible pathways for an electromagnetic wave to interact with the volume to be occupied by the defect. 

Furthermore, the tensor $\Gamma^{(d)}$ yields the exact scattered field, which follows from substitution of \eqref{eq:resolvent-IE-1} into  \eqref{eq:E-formal-1}, and removing the zeroth-order term,
\begin{equation}
    \Delta E_i(\boldsymbol{r}) = 
    k^2\int d\boldsymbol{r}_1\;
    G^{(0)}_{ij}(\boldsymbol{r},\boldsymbol{r}_1)
    \int \mathrm{d}^3\boldsymbol{r}_2\;
    \Gamma^{(d)}_{jj'}(\boldsymbol{r}_1,\boldsymbol{r}_2)
    E^{(0)}_{j'}(\boldsymbol{r}_2).\label{eq:scattered-E-1}
\end{equation}
This equation describes a perturbed polarization propagating through the background medium and generating the entire scattered field. Equation \eqref{eq:gamma-IE-1} then shows that,
\begin{eqnarray}
P^{(d)}_i(\boldsymbol{r}) &=& 
\epsilon_0\int \mathrm{d}^3\boldsymbol{r}'\;\; \Gamma^{(d)}_{ij}(\boldsymbol{r},\boldsymbol{r}') 
E^{(0)}_j(\boldsymbol{r}')\\
&=&\epsilon_0\chi^{(d)}_{ij}(\boldsymbol{r})E^{(0)}_j(\boldsymbol{r}) + 
\epsilon_0 \chi^{(d)}_{ij}(\boldsymbol{r})\Delta E_j(\boldsymbol{r}).\label{eq:Pd-Gammad-E0}
\end{eqnarray}
Thus $\boldsymbol{P}^{(d)}$ captures the complete change in the polarization within the volume of the defect. The first term is simply the first order Born approximation. The second term in \eqref{eq:Pd-Gammad-E0} represents the self-interaction in which the scattered field interacts with the perturbation that generated it. 

Let us now return to a detailed discussion of \eqref{eq:resolvent-IE-1} and \eqref{eq:gamma-IE-1} and the approximations that follow from them. While these equations are equivalent in their exact non-local form represented by infinite series, they generate different results for the local approximations. Following Habashy \textit{et al.}\cite{Habashy1993}, a local approximation for a kernel $g(x,x')$ that is singular at $x=x'$ is generated by writing it as,
\begin{equation}
    \int dx' g(x,x') f(x') = \left[\int dx' g(x,x')\right]f(x) + \int dx' g(x,x') [f(x')-f(x)]. 
\end{equation}
On the right hand side, each expression involves only non-singular terms as the factor multiplying $g(x,x')$ is zero in the second term. While this equation is exact, a local approximation is made by dropping the second term, which is justified when the function $f(x)$ is smooth near the singularity of $g$ and the kernel $g(x,x')$ decays rapidly with $x-x'$. For simple kernels, such as Green's function in free space, the dependence on $x-x'$ is known and can be used to assess error in the approximation. Here $g$ corresponds to the Green function $\boldsymbol{G}^{(0)}$ which does not have an analytical form, but it must possess an integrable singularity inside a volume element of single dielectric constant due to \eqref{eq:zeroth-GF-waveq}. Thus we expect the approximation to hold, and its validity is assumed not proved. 

When the above splitting of the kernel is applied to \eqref{eq:resolvent-IE-1} and \eqref{eq:gamma-IE-1}, 
\begin{eqnarray}
    \overleftrightarrow{\mathcal{R}}^{(d)}(\boldsymbol{r},\boldsymbol{r}')&=& 
    \overleftrightarrow{\mathcal{L}}(\boldsymbol{r})\delta(\boldr-\boldr')+
    k^2\overleftrightarrow{\mathcal{L}}(\boldsymbol{r})
    \Delta \overleftrightarrow{\mathcal{R}}(\boldsymbol{r},\boldsymbol{r}')\label{eq:resolvent-eq-renorm}\\
    \overleftrightarrow{\Gamma}^{(d)}(\boldsymbol{r},\boldsymbol{r}')&=& 
    \overleftrightarrow{\Lambda}(\boldsymbol{r})\delta(\boldr-\boldr')+
    k^2\overleftrightarrow{\Lambda}(\boldsymbol{r})
    \Delta \overleftrightarrow{\Gamma}(\boldsymbol{r},\boldsymbol{r}').\label{eq:Gamma-eq-renorm}
\end{eqnarray}
In the above equations, I have defined the local operators,
\begin{eqnarray}
\overleftrightarrow{\mathcal{L}}(\boldsymbol{r}) & = & 
\left[\overleftrightarrow{I}-k^2\int_{\defectvol} \mathrm{d}^3\boldsymbol{r}''
    \overleftrightarrow{G}^{(0)}(\boldsymbol{r},\boldsymbol{r}'')
     \overleftrightarrow{\chi}^{(d)}(\boldsymbol{r}'')\right]^{-1},\label{eq:local-field-R}\\
\overleftrightarrow{\Lambda}(\boldsymbol{r}) & = & 
\left[
\overleftrightarrow{I}-
k^2\overleftrightarrow{\chi}^{(d)}(\boldsymbol{r})
\int_{\defectvol} \mathrm{d}^3\boldsymbol{r}''
\overleftrightarrow{G}^{(0)}(\boldsymbol{r},\boldsymbol{r}'')\right]^{-1}
\overleftrightarrow{\chi}^{(d)}(\boldsymbol{r}),\label{eq:local-field-Gamma}
\end{eqnarray}
and the non-local corrections
\begin{eqnarray}
\Delta \overleftrightarrow{\mathcal{R}}(\boldsymbol{r},\boldsymbol{r}')
    &=&\int_{\defectvol} \mathrm{d}^3\boldsymbol{r}''
    \overleftrightarrow{G}^{(0)}(\boldsymbol{r},\boldsymbol{r}'')
     \overleftrightarrow{\chi}^{(d)}(\boldsymbol{r}'')
     \left[ 
     \overleftrightarrow{\mathcal{R}}(\boldsymbol{r}'',\boldsymbol{r}') -  
     \overleftrightarrow{\mathcal{R}}(\boldsymbol{r},\boldsymbol{r}') \right] \label{eq:nonlocalR}
    \\   
\Delta \overleftrightarrow{\Gamma}(\boldsymbol{r},\boldsymbol{r}') &=&
    \int_{\defectvol} \mathrm{d}^3\boldsymbol{r}''
    \overleftrightarrow{G}^{(0)}(\boldsymbol{r},\boldsymbol{r}'')
    \left[
    \overleftrightarrow{\Gamma}(\boldsymbol{r}'',\boldsymbol{r}')-
    \overleftrightarrow{\Gamma}(\boldsymbol{r},\boldsymbol{r}')
    \right].
\end{eqnarray}
Dropping the second term in equation \eqref{eq:resolvent-eq-renorm} is the approximation developed by Torres-verdin and Habashy. The tensor $\overleftrightarrow{\mathcal{L}}$ is often called the \emph{depolarization tensor} in the literature\cite{kristensson2016scattering}. The depolarization tensor transforms the background (bare) field to the higher order corrections of the total field. On the other hand, $\localLambda$ transforms the Born approximation of the defect polarization into higher order corrections for the induced polarization. The two are equivalent when $\overleftrightarrow{\chi}^{(d)}(\boldsymbol{r})$ is uniform and commutes with the background Green function. Thus one expects to see a difference between these approximations for anisotropic media and with non-uniform defect dielectric constant. 

Armed with the formalism that connects the local approximations to the exact equations through both the field and the polarization as the fundamental objects to approximate, an investigation of the relative merits of these choices will be presented in future publications. In the remaining paper, my goal is to develop and test numerical methods to estimate the  tensors above without explicit calculations of Green functions.

\subsection{Estimation of Depolarization Tensor}\label{sec:parameter estimation}

For clarity of discussion in this section, let us first define a family of source polarization functions indexed by $\alpha$. A source $\boldsymbol{P}^{s}_\alpha(\boldsymbol{r})$ drives the system to excite the field $\boldsymbol{E}^{(0)}_\alpha(\boldsymbol{r})$ as the solution to the background field, and $\boldsymbol{E}_\alpha(\boldsymbol{r})$ as the total field in the presence of the defect. As the sources are placed in the upper and lower half spaces infinitely far away from the structure,  the fields $\boldsymbol{E}_\alpha(\boldsymbol{r})$ asymptotically approach the eigenmodes of the upper and or lower half spaces. These modes satisfy the incoming and outgoing wave boundary conditions. Thus $\boldsymbol{P}^{s}_\alpha(\boldsymbol{r})$ can be viewed alternatively as an \emph{eigenmode} source, which I describe simply by $\alpha$. In the case of uniform upper and lower half planes, these modes are always upward and downward propagating plane waves. In the following I will use $\bar{\alpha}$ to indicate the mode counter-propagating to the one described by $\alpha$.

The basic principle of this approach is as follows. First compute the background fields  $\boldsymbol{E}^{(0)}_\alpha$, and the total fields and $\boldsymbol{E}_\alpha$, for several $\alpha$. Using these fields as the data, and recognizing that the tensors introduced above do not depend on $\alpha$, the strategy is to solve the inverse problem for $\Lambda_{ij}(\boldsymbol{r})$ and $\Delta \Gamma_{ij}(\boldsymbol{r},\boldsymbol{r}')$. Since these two quantities cannot be solved exactly over the entire continuous space without a large number of such computations, one determines them only inside $V^d$, which is the defect volume. Furthermore, one constructs them only to a limited resolution by  expanding them into a finite set of real-valued basis functions (with complex-valued coefficients),
\begin{eqnarray}
\Lambda_{ij}(\boldsymbol{r}) &\approx& \sum_{n=1}^{N_b} \Lambda_{ij,n} \varphi_n(\boldsymbol{r}),\label{eq:Lambda-basis-expansion}\\
\Delta\Gamma_{ij}(\boldsymbol{r},\boldsymbol{r}') &\approx& \sum_{n,m=1}^{N_b} \Gamma_{ij,nm} \varphi_n(\boldsymbol{r})\varphi_m(\boldsymbol{r}').\label{eq:dgamma-basis-expansion}
\end{eqnarray}

The expansion coefficients in \eqref{eq:Lambda-basis-expansion} and \eqref{eq:dgamma-basis-expansion} are determined by minimizing an objective function involving various physical quantities of interest. In this work, it is the scattered field in the far zone, or more precisely, the scattering matrix elements between incidence and radiation states into the upper half space of the structure that is important. However, the formalism developed below is not limited to this, and I also include the energy stored in polarizing the defect volume as part of the general case of interest. The objective function for the inverse problem is thus stated by minimizing the difference between these quantities computed directly from $\boldsymbol{E}^{(0)}_\alpha$on the one hand, and via equations \eqref{eq:E-formal-1}, \eqref{eq:scattered-E-1}, \eqref{eq:resolvent-eq-renorm}-\eqref{eq:nonlocalR} on the other. I will denote the latter with a tilde to indicate that it is an approximate quantity. 

I now turn to constructing the objective function to solve the inverse problem. From the scattered field equation \eqref{eq:scattered-E-1}, and the definition of defect polarization, \eqref{eq:Pd-Gammad-E0}, I define a matrix element between state $\alpha$ and $\beta$ as,
\begin{eqnarray}
   U_{\alpha \beta} &=& \frac{k^2}{\epsilon_0}\sum_{i,j\in x,y,z} \int \mathrm{d}^3\boldsymbol{r} \int_{V^d} \mathrm{d}^3\boldsymbol{r}' P^s_{\bar{\alpha},i}(\boldsymbol{r})G^{(0)}_{ij}(\boldsymbol{r},\boldsymbol{r}') P^{(d)}_{\beta,j}(\boldsymbol{r}').\label{eq:U Ps G0 Pd}
\end{eqnarray}
The matrix element has dimensions of energy. Recall that $\bar{\alpha}$ corresponds to the field counter-propagating to $\alpha$, which for plane wave states would be $(-\boldsymbol{k}_\alpha,\hat{e}_\alpha)$. Introduction of this extra notation is necessary to naturally account for the Fourier transform relation that maps the real-space scattered field to its Fourier component that is the actual S-matrix element.

The systems of interest in this work do not have quasi-static magnetic fields and are thus time-reversal invariant. Under this condition, Onsager reciprocity \cite{lax} holds for all linear response tensors of the electric permittivity, and thus the Lorentz reciprocity also holds \cite{jackson1999classical} for the electromagnetic fields. By applying the reciprocity theorem \cite{jackson1999classical}, the background Green function can be eliminated by integrating over $\boldsymbol{r}$ and using the definition \eqref{eq:E0-G-P}, which yields the expression, 
\begin{eqnarray}
   U_{\alpha \beta} &=& \sum_{i \in x,y,z} \int_{V^d} \mathrm{d}^3\boldsymbol{r} \boldsymbol{E}^{(0)}_{\bar{\alpha}}(\boldsymbol{r}) \cdot \boldsymbol{P}^{(d)}_{\beta}(\boldsymbol{r}).\label{eq:Uexact}
\end{eqnarray}
In the case of plane wave sources for which $\alpha$ specifies the wavevector and the polarization state in the plane orthogonal to the wavevector, $(\boldsymbol{k}_\alpha,\hat{e}_\alpha)$, the scattering matrix of the defect, in terms of $U_{\alpha\beta}$ is,
\begin{equation}
\Delta S(\boldsymbol{k}_\alpha,\hat{e}_\alpha;\boldsymbol{k}_\beta,\hat{e}_\beta) = \frac{ik^2}{2 \epsilon_0 E_0 k_z} U_{\alpha\beta}.\label{eq:s-matrix-Uab}
\end{equation}
Here I have defined $E_0$ as numerically a unit amplitude of the incident plane wave but indicated explicitly since it carries the units of the electric field. Since $U_{\alpha\beta}$ has dimensions of energy, it follows that $\Delta S$ has dimensions of Volts $\times$ length, as it must be for the Fourier transformed electric field. The pre-factors follow by writing $\delta(z-z_s)e^{i\boldsymbol{\kappa}\cdot\boldsymbol{r}_\perp}$ in terms of the expression of $P^s$ given in Appendix \ref{app:sources}. 

The matrix defined in \eqref{eq:Uexact} is with the ab-initio total fields, that are to be computed "exactly" within a tolerable numerical error. The approximation of this matrix from only the background fields and the tensors follows from \eqref{eq:resolvent-eq-renorm}, and is,
\begin{eqnarray}
    \tilde{U}_{\alpha \beta} &=& \sum_{ij} \int_{V^d} \mathrm{d}^3\boldsymbol{r} E^{(0)}_{\bar{\alpha},i}(\boldsymbol{r})  \Lambda_{ij}(\boldsymbol{r}) %
    E^{(0)}_{\beta,j}(\boldsymbol{r}) + \nonumber \\
&& \sum_{iji'j'} k^2
\int_{V^d} \mathrm{d}^3\boldsymbol{r}  E^{(0)}_{\bar{\alpha},i}(\boldsymbol{r}) 
\Lambda_{ij'}(\boldsymbol{r}) %
    \int_{V^d} \mathrm{d}^3\boldsymbol{r}' \Delta\Gamma^{(d)}_{j'j}(\boldsymbol{r},\boldsymbol{r}') E^{(0)}_{\beta,j}(\boldsymbol{r}'). 
\label{eq:Uab_equation2} %
\end{eqnarray}
The sums are over Cartesian indices of the fields. The restriction of the domain for $\boldsymbol{r}'$  integration in the second term follows from the restriction of $\boldsymbol{r}$ due to the presence of $\chi^{(d)}(\boldsymbol{r})$ and the integral equation \eqref{eq:resolvent-IE-1}. To  facilitate a compact formulation, I define the coefficient matrices,
\begin{eqnarray}
 E^{(0)}_{\beta;i,n} &=& \int_{V^d} \mathrm{d}^3\boldsymbol{r} \varphi_{n}(\boldsymbol{r}) E^{(0)}_{\beta,i}(\boldsymbol{r}), \label{eq:E-phi} \\ 
 \mathcal{T}_{\alpha\beta;ij,n} &=&
 \int_{V^d} \mathrm{d}^3\boldsymbol{r} E^{(0)}_{\bar{\alpha},i}(\boldsymbol{r})\varphi_{n}(\boldsymbol{r})E^{(0)}_{\beta,j}(\boldsymbol{r}),\label{eq:T1} \\
\mathcal{T}_{\alpha\beta;ij,p,nm} &=& k^2 \int_{V^d}  \mathrm{d}^3\boldsymbol{r} E^{(0)}_{\bar{\alpha},i}(\boldsymbol{r})  
\varphi_{p}(\boldsymbol{r}) \varphi_{n}(\boldsymbol{r})  E^{(0)}_{\beta;jm}.\label{eq:T2}
\end{eqnarray}
With these definitions, one obtains the following algebraic expression for the matrix elements,
\begin{eqnarray}
\tilde{U}_{\alpha \beta} &=& \sum_{ijn}\mathcal{T}_{\alpha\beta;ij,n} \Lambda_{ij,n} + \sum_{ii'jpnm}\mathcal{T}_{\alpha\beta;ij,p,nm} \Lambda_{ii'p} %
\Delta\Gamma_{i'j,nm}\;.\label{eq:U-tilde}
\end{eqnarray}
One may also develop an additional set of constraints by integrating the energy density in the polarization induced inside $V^d$  \cite{jackson1999classical},
\begin{equation}
     W_{\alpha} = \sum_i \frac{1}{2}\int_{V^d}  E^*_{\alpha,i}(\boldsymbol{r})P_{\alpha,i}(\boldsymbol{r}) \mathrm{d}^3\boldsymbol{r}.
\end{equation}
The corresponding $\tilde{W}_{\alpha}$ based on the approximation of the total field via resolvent operator, is
\begin{eqnarray}
\tilde{W}_{\alpha } &=& \sum_{ijn}\mathcal{W}_{\alpha;ij,n} \Lambda_{ij,n} + \sum_{ii'jpnm}\mathcal{W}_{\alpha;ij,p,nm} \Lambda_{ii'p} %
\Delta\Gamma_{i'j,nm}\;, 
\end{eqnarray}
where the coefficient matrices are defined as shown below,
\begin{eqnarray}
 \mathcal{W}_{\alpha;ij,n} &=&\frac{1}{2} \int_{V^d} \mathrm{d}^3\boldsymbol{r} E^{(0)*}_{\alpha,i}(\boldsymbol{r})\varphi_{n}(\boldsymbol{r})E^{(0)}_{\beta,j}(\boldsymbol{r}), \label{eq:W1}\\
\mathcal{W}_{\alpha;ij,p,nm} &=&\frac{k^2}{2}\int_{V^d} \mathrm{d}^3\boldsymbol{r} E^{(0)*}_{\alpha,i}(\boldsymbol{r})
\varphi_{p}(\boldsymbol{r}) \varphi_{n}(\boldsymbol{r}) E^{(0)}_{\alpha;jm}.\label{eq:W2}
\end{eqnarray}

Finally, I can state the objective function in terms of the scattered field components and the energy of interaction as,
\begin{equation}
    Q(\eta, \Lambda_{ijn},  \Delta\Gamma_{ijnm}) = \sum_{\alpha,\beta} |U_{\alpha,\beta}-\tilde{U}_{\alpha,\beta}|^2 + \eta \sum_\alpha |W_{\alpha}-\tilde{W}_{\alpha}|^2,\label{eq:cost-function}
\end{equation}
where $\eta$ is the relative weight placed on matching the energy stored. I consider the simplest cases where $\eta=0$ so that only the diffraction orders are computed accurately in the next section. I will determine the impact of increasing $\eta$ in a future publication. The objective function is minimized to obtain the solution to the inverse problem in terms of the expansion coefficients $\Lambda_{ij,n}$ and $\Delta\Gamma_{ij,nm}$. When the non-local term $\Delta\Gamma_{ij,nm}$ can be ignored, the problem reduces to linear least squares estimation of $\Lambda_{ij,n}$. Furthermore, the solution must be assessed based on its generalization error to data not included in the process of the finding the solution, or in other words the \emph{training set}. This is necessary to ensure that these tensors are independent of the sources.

Before leaving this section, I remark that increasing the number of basis functions, $N_b$, increases the accuracy for the physical quantities of interest. To avoid over-fitting, it must also be accompanied by the corresponding increase in data, and therefore the ab-initio computations of the fields. Since the cost of the latter is high, $N_b$ is limited by these computations, and hence the main premise of the paper. Numerical tests on a variety of structures suggest that $N_b \leq 3$ per dielectric material in the defect volume suffices for a majority of the applications. These bases correspond to transforming the polarization up to dipole and quadrupole orders. This may not be the correct transformation point-by-point, but it yields the correct integral quantities such as diffraction orders and interaction energy.

\subsection{Numerical Implementation}\label{sec:numerical implementation}
I now describe a specific numerical implementation I have found to be effective in applications of this work. I assume that the electric fields in the equations of the previous section are stored as samples on a set of points filling up the space in $V^d$. I decompose the domain into a set of Voronoi cells, $\{v_I\}$ such that each has a single value of $\chi^{(d)}$ and a single sample $\boldsymbol{r}_I$, and then define two types of basis-function matrices over the Voronoi cell vertices, 
\begin{eqnarray}
\Phi_n(I) &=& \int_{v_I} \mathrm{d}^3\boldsymbol{r} \varphi_{n}(\boldsymbol{r}) ,\label{eq:Phi_nIK} \\
\Phi_{nm}(I) &=& \int_{v_I} \mathrm{d}^3\boldsymbol{r} \varphi_{n}(\boldsymbol{r})\varphi_{m}(\boldsymbol{r}).\label{eq:Phi_nmIK}
\end{eqnarray}
The integrals in equations \eqref{eq:E-phi}-\eqref{eq:T2} now take the form
\begin{eqnarray}
 E^{(0)}_{\beta;jn} &=& \sum_I E^{(0)}_{\beta,j}(\boldsymbol{r}_{I}) \Phi_{n}(I) , \label{eq:E-phi-voronoi} \\ 
 \mathcal{T}_{\alpha\beta;ij,n} &=& \sum_I
 E^{(0)}_{\bar{\alpha},i}(\boldsymbol{r}_{I})E^{(0)}_{\beta,j}(\boldsymbol{r}_{I}) \Phi_{n}(I),\label{eq:T1-voronoi} \\
\mathcal{T}_{\alpha\beta;ij,p,nm} &=& \left[k^2 \sum_I E^{(0)}_{\bar{\alpha},i}(\boldsymbol{r}_I) \Phi_{pn}(I) \right] E^{(0)}_{\beta;jm}.\label{eq:T2-voronoi}
\end{eqnarray}
Similarly, the coefficient matrices defined in \eqref{eq:W1} and \eqref{eq:W2} can be represented as summations as follows,
\begin{eqnarray}
 \mathcal{W}_{\alpha;ij,n} &=&\frac{1}{2} \sum_I  E^{(0)*}_{\alpha,i}(\boldsymbol{r}_{I})E^{(0)}_{\beta,j}(\boldsymbol{r}_{I}) \Phi_n(I), \\
\mathcal{W}_{\alpha;ij,p,nm} &=&\left[\frac{k^2}{2} \sum_I 
E^{(0)*}_{\alpha,i}(\boldsymbol{r}_{I}) \Phi_{pn}(I)
 \right]E^{(0)}_{\alpha;jm}.\label{eq:W-voronoi}
\end{eqnarray}
The basis functions are arbitrary. In the present work, I chose from two types of functions. 

The first set of functions, which suffice for small defects, are $M$ piecewise constant function where $M$ is number of unique materials comprising the defect volume $V^d$. Therefore I write the uniform $\chi^{(d)}(\boldsymbol{r})$ as $\chi^{(d)}_n$ inside each sub-domain $J$,  
\begin{equation}
\varphi_{n}(\boldsymbol{r}) = \Theta\left(\chi^{(d)}(\boldsymbol{r})-\chi^{(d)}_n\right),\;\;\ n=1,\ldots,M,\label{eq:cg-basis}
\end{equation}
where $\Theta$ is the indicator function with the value 1 when its argument is zero, and 0 for all other complex values. These functions represent a rigid transformation of the polarization in each dielectric sub-domain of the defect. For defects where some of the sub-domains are large enough that a rigid transformation cannot represent the change in polarization accurately, I introduce additional functions in the form of so-called Cartesian Gaussian Type Orbitals (CGTOs) \cite{boys1950electronic}. These functions construct higher order multipoles around the center of the sub-domain, or multiple centers in larger domains. There are 3 CGTOs representing a set of reducible functions for multipole order of $l=1$.
\begin{eqnarray}
\varphi_{1,2,3}(\boldsymbol{r}) &=& (xe^{-\zeta r^2},ye^{-\zeta r^2},ze^{-\zeta r^2}).
\end{eqnarray}

The discussion in this and the preceding section combines statistical estimation with deterministic solutions to Maxwell equations. In particular, the source-dependent quantities in the model equation \eqref{eq:U-tilde} are computed deterministically, while the unknowns $\Lambda_{ij}$ are solved via estimation since a fully invertible system is not guaranteed. The discussion below thus relies on some basic terminology of the statistical estimation theory and practice. Readers that may be unfamiliar with the terms may refer to Appendix \ref{sec:terminology}.

\section{Results of Numerical Calculations}\label{sec:results}

\begin{figure}[bh]
\centering
\begin{subfigure}[b]{0.4\textwidth}
\includegraphics[height=2.in]{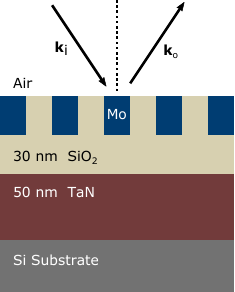}
\subcaption{}
\end{subfigure}
\begin{subfigure}[b]{0.4\textwidth}
\centering
\includegraphics[height=1.5in]{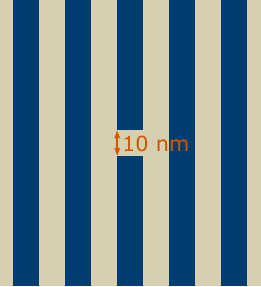}
\subcaption{}
\end{subfigure}

\caption{(a) Schematic vertical Cross section of the line-space geometry used in the calculations in this section, and (b) schematic top view of the pattern with a 10 nm open circuit defect. The pitch varies between 20 nm and 40 nm, while the material and thickness of each layer is fixed among all calculations. The pattern layer at top is also 30 nm thick. Although Mo is not a metal used in the present technology node, I have simply used it as an example as it is poised to be a metal of choice in the future\cite{imec}.}

\label{fig:Geometry}

\end{figure}

Let us now test the above theoretical formulation by applying it to the calculation of scattering matrices of an isolated defect in a semi-infinite background structure. Figure \ref{fig:Geometry} shows diagrams of the line-space system used as a test case. These systems represent many of the back-end-of-line layers in the semiconductor chip manufacturing as they form intermediate layers of interconnections from external electrical circuitry to the nanometer scale transistors. These layers are characterized by metal lines with an insulator material in between, which is typically a dielectric such as SiO$_2$ or its low density forms. The defects in these structures that can electrically disable the device are the breaks in a metal line or a metal bridge across an oxide region. These breaks and bridges typically range from 10 nm to 40 nm in size in the present technology node. I describe this structure in more detail in the Supplemental Materials section. There, I also discuss our computation of the electric fields using the FDTD package MEEP.

As the testing methodology, I performed 3 different calculations of the scattering matrix of the defect. In the first calculation, I performed a direct calculation of the scattering matrix using \eqref{eq:Uexact} and \eqref{eq:s-matrix-Uab}. The matrix elements are computed as $\Delta S(\boldsymbol{k},\hat{e};\boldsymbol{k}',\hat{e}')$, where $\hat{e},\hat{e}'$ are the "s" and "p" polarization directions $\hat{s}=\hat{\kappa}\times\hat{z}$ and $\hat{p}=\hat{s}\times\hat{k}/|\boldsymbol{k}|$. The test set was generated with up to 90 plane wave directions for incident and outgoing set. This yielded a test dataset of more than 5000 non-zero matrix elements to test the predictions of the theory presented above. 

In the second calculation, I performed the standard Born approximation to predict these test matrix elements. This calculation was performed as a base case to demonstrate clearly the improvement our technique makes.

In the third calculation, I compute the matrix elements using the theory presented above, which I call the \emph{tensor approximation} here. I formed a training set of up to 13 k-points and restricted \emph{both} the incident $\boldsymbol{k}'$ and the outgoing $\boldsymbol{k}$ wave vectors to this subset. Generation of this training set thus requires only 13 \emph{ab-initio} calculations. I then used this subset to compute the tensor $\Lambda$ from \eqref{eq:Uab_equation2} and the cost function \eqref{eq:cost-function} in which I set $\eta=0$. The overlap integrals between the fields are computed following the scheme described in Sec. \ref{sec:numerical implementation}. 

When applying the CGTO basis set, I computed estimates of $\Lambda$ using 4 combinations of the basis functions $\{1\}$, $\{1,x,y\}$, and $\{1,x,y,z\}$. Since the overlap integrals are stored over the largest basis set considered, these extra computations incur negligible cost. I selected the basis combination that best fit the ground truth in each case. This represents a type of model selection, which I implement in practice by splitting the calculated fields into a training and validation set. The latter is then used to evaluate the generalization error. I remark that it is not the number of incidence directions, but the total number of matrix elements that is important in the fitting. I elaborate this point, and how it relates to this calculation in depth in the next section. I now turn to presenting the results of the three calculations discussed above.

\begin{figure}[bh]
\centering

\begin{subfigure}[c]{0.3\textwidth}
\centering
\includegraphics[height=1.in]{LS20_10.pdf}
\subcaption{Pitch = 20 nm}
\end{subfigure}
\begin{subfigure}[c]{0.66\textwidth}
\centering
\includegraphics[height=1.5in]{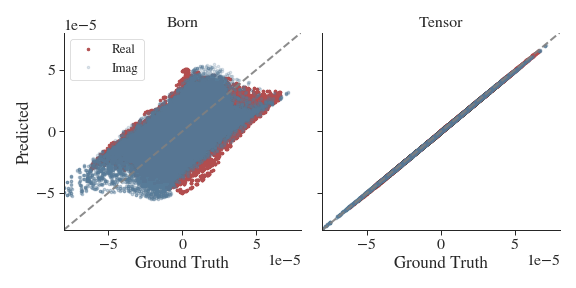}
\end{subfigure}

\begin{subfigure}[c]{0.3\textwidth}
\centering
\includegraphics[height=1.in]{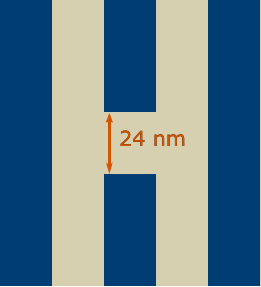}
\subcaption{Pitch = 40 nm}
\end{subfigure}
\begin{subfigure}[c]{0.66\textwidth}
\centering
\includegraphics[height=1.5in]{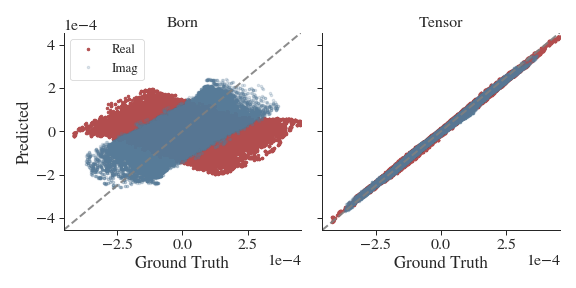}
\end{subfigure}

\begin{subfigure}[c]{0.3\textwidth}
\centering
\includegraphics[height=1.in]{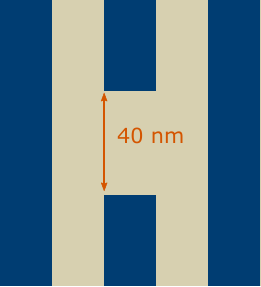} 
\subcaption{Pitch = 40 nm}
\end{subfigure}
\begin{subfigure}[c]{0.66\textwidth}
\centering
\includegraphics[height=1.5in]{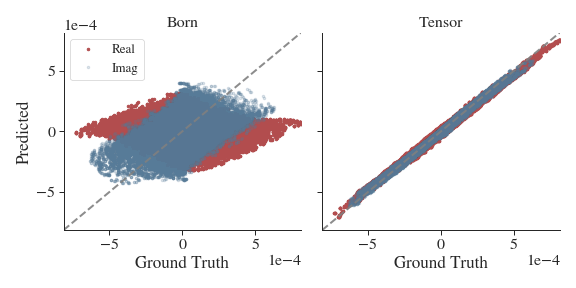}
\end{subfigure}
\caption{Correlation plots of S-matrix components ($\lambda = 200$ nm) between the \emph{ab-initio} calculation and those predicted by Born approximation and its extension described in this work. The calculations are for the line space geometry with a dielectric (light yellow) break in the metal line (dark blue), and the pitch and the defect sizes as indicated. The metal lines are assigned the dielectric constant of Mo while that for the dielectric is SiO$_2$ sourced from Palik \emph{et al.}. All three cases have identical layers in the $z$-direction as shown in Fig.\ref{fig:Geometry} . The correlation coefficient $R^2$ (see Appendix \ref{sec:terminology}) in each case is, (a) Born: 0.339 , Tensor: 1.000, (b) Born: 0.059, Tensor: 0.997, (c) Born: 0.035,  Tensor: 0.994.}

\label{fig:Mo200}
\end{figure}

In Fig. \ref{fig:Mo200}, I plot the real and imaginary parts of the matrix elements from the Born approximation and from the tensor approximation against their counterparts in ground truth. I show the results for 10 nm, 24 nm, and 40 nm defects in 20 nm and 40 nm pitch line-space system as depicted in the leftmost column of the figure. It is clear from these plots that the Born approximation fails to predict both the magnitude and the phase of the scattering matrix. On the other hand, the extension presented above, which I call the Tensor approximation shows an almost perfect correlation even for defect size of 20$\times$40 nm$^2$, which are much larger than those of interest in the future technology nodes. 

The comparison between the predictions and the ground truth is performed via correlation plot, which may seem unconventional. These plots are chosen to  show the full extent of the values of the S-matrix components, and the deviation of the predicted results from them. Summary of the differences is provided in Table \ref{tab:error} by normalizing them to a measure of the magnitude of the matrix elements. The left half the table shows the normalized error in S-matrix values for the three different structures shown in Fig. \ref{fig:Geometry}, and defined as
\begin{equation}
\epsilon(\tilde{U}) = \frac{\sqrt{\sum_{\alpha\beta} |\tilde{U}_{\alpha\beta} - U_{\alpha\beta}|^2 } }{\sqrt{\sum_{\alpha\beta} |U_{\alpha\beta}|^2 } }. \label{eq:rel-err-S}
\end{equation}
Recall that $U_{\alpha\beta}$ is the numerically exact S-matrix computed directly from the solutions of the Maxwell equations. The error is normalized to account for the changes in the magnitude of the matrix elements. Relative error of individual elements is not informative since some of the matrix elements are nearly zero and therefore contribute little to the final results while they dominate the relative errors. Note that neither the correlation nor the arithmetic differences are directly related to the final physical quantities of interest. 

In the present case, the final quantities are the intensity images obtained from the Abbe image theory\cite{kitkwong}. Thus I computed incoherent images under two conditions: (i) very low numerical aperture (NA)\footnote{Numerical aperture, $NA=\sin{\theta}$ where $\theta$ is the angle of a wave vector of a plane wave with respect to the surface normal.}, and (ii) annular illumination at high-NA. The intensity in the image plane is computed from the Abbe theory \cite{kitkwong}. I compute one image with the base structure, which is unresolved and therefore a uniform intensity, and subtract it from the image with the defect present. These difference images allow sub-wavelength defects to be detected as anomalies. The signal strength is the maximum absolute value of the difference intensity.\footnote{In practice, this technique requires exquisite control of the imaging process and advanced algorithms for hardware and software based alignment.}. The image calculation is described mathematically in Appendix \ref{sec:ImageCalc}. Next, I show that computing these difference images using the Tensor approximation incurs little error with respect to the ground truth.

\begin{figure}[bh]
\centering

\begin{subfigure}[c]{0.45\textwidth}
\centering
\includegraphics[height=1.in]{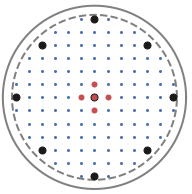}
\end{subfigure}\hfill
\begin{subfigure}[c]{0.45\textwidth}
\centering
\includegraphics[height=1.in]{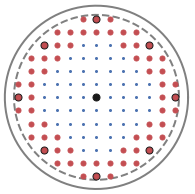}
\end{subfigure}

\begin{subfigure}[c]{0.45\textwidth}
\centering
\includegraphics[height=1.5in]{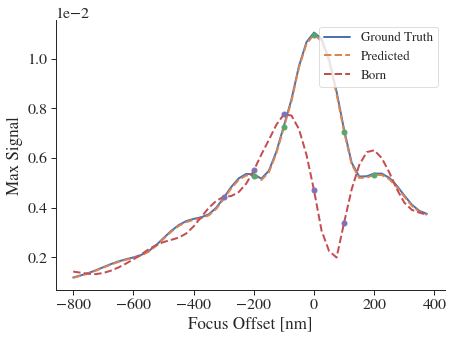}
\subcaption{}
\end{subfigure}\hfill
\begin{subfigure}[c]{0.45\textwidth}
\centering
\includegraphics[height=1.5in]{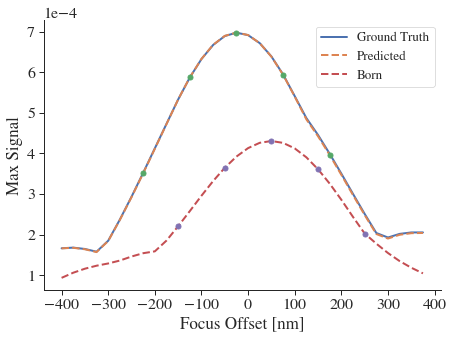}
\subcaption{}
\end{subfigure}

\begin{subfigure}[c]{0.45\textwidth}
\centering
\includegraphics[height=1.5in]{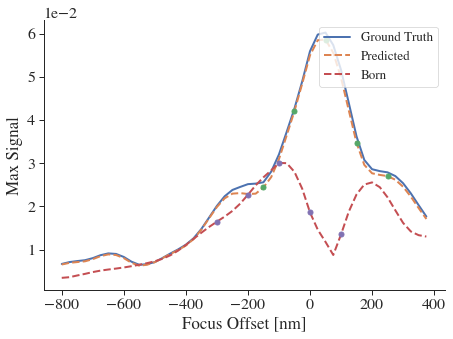}
\subcaption{}
\end{subfigure}\hfill
\begin{subfigure}[c]{0.45\textwidth}
\centering
\includegraphics[height=1.5in]{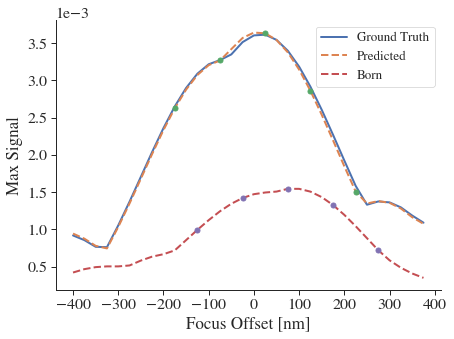}
\subcaption{}
\end{subfigure}

\begin{subfigure}[c]{0.45\textwidth}
\centering
\includegraphics[height=1.5in]{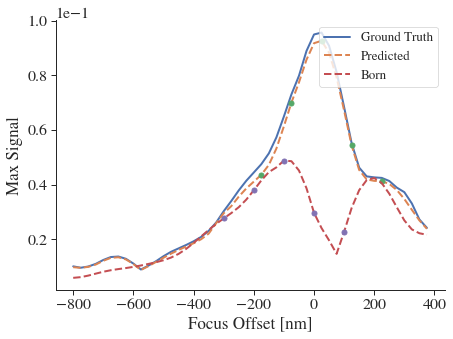}
\subcaption{}
\end{subfigure}\hfill
\begin{subfigure}[c]{0.45\textwidth}
\centering
\includegraphics[height=1.5in]{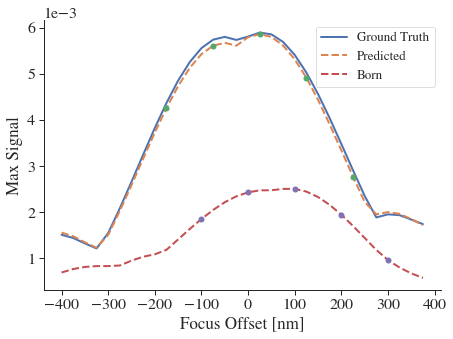}
\subcaption{}
\end{subfigure}

\caption{Focus dependence of the signal for the correlation plots shown in Fig. \ref{fig:Mo200}. The dots along the curves mark the offsets where the actual images are shown in figures \ref{fig:Mo200-thru-focus1}-\ref{fig:Mo200-thru-focus3}. The top row show the grid of in plane wavevectors $\boldsymbol{\kappa}$ that was sampled for generating the full test set and plotted in dimensionless units $\boldsymbol{\kappa}/k$ where the outer solid circle represent the grazing incidence angle of 90 degrees to the surface normal. The red dots indicate the plane waves selected for illumination to form the image. The black dots show the incidence directions in the training set, and therefore the only points where \textit{ab-initio} calculations must be performed for computing the tensor approximation.} 
\label{fig:Mo200-focus-scans}
\end{figure}

In Fig. \ref{fig:Mo200-focus-scans}, I show the signal strength as a function of focus offset for all the cases presented above, and for the aforementioned two different illumination patterns, shown also at the top of the figure. Since these images are bright-field images, they include the interference between the defect and the background field. We see that the focus dependence of the signal from Tensor approximation follows closely the ground truth over a focus range of 6 wavelengths. Since the approximation is correct both in the magnitude and the phase of the matrix elements, the focus dependence is expected to be correct. On the other hand, the Born approximation due to its incorrect phase prediction will not produce the correct prediction for the focus behavior, even if one scales it to correct the magnitude. 

\begin{figure}[bh]
\centering
\includegraphics[height=2.in]{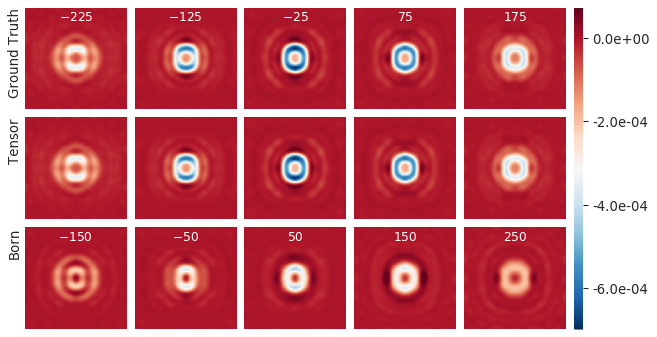}
\caption{Images at focus offsets shown as dots in Fig. \ref{fig:Mo200-focus-scans}(b).}
\label{fig:Mo200-thru-focus1}
\end{figure}

\begin{figure}[bh]
\centering
\includegraphics[height=2.in]{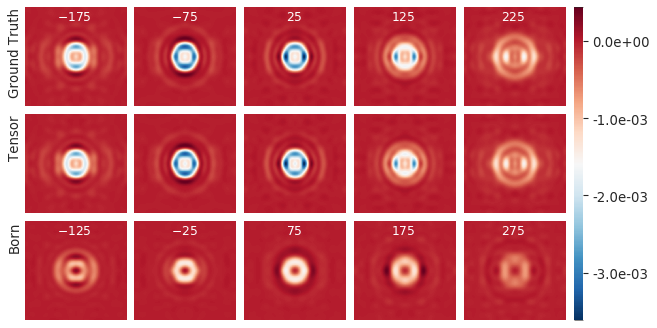}
\caption{Images at focus offsets shown as dots in Fig. \ref{fig:Mo200-focus-scans}(d).}
\label{fig:Mo200-thru-focus2}
\end{figure}

\begin{figure}[bh]
\centering
\includegraphics[height=2.in]{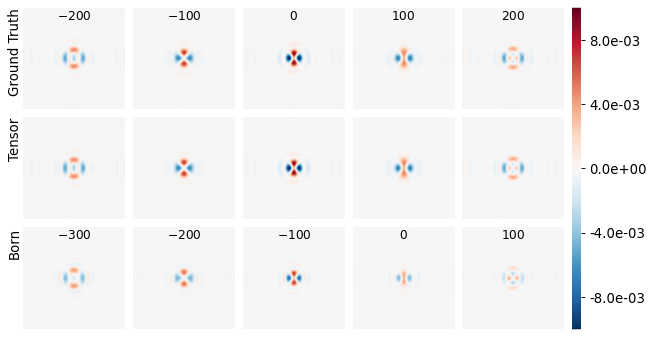}
\caption{Images at focus offsets shown as dots in Fig. \ref{fig:Mo200-focus-scans}(e).}
\label{fig:Mo200-thru-focus3}
\end{figure}

In figure \ref{fig:Mo200-thru-focus1}, I show the actual images at 5 focus offsets. The focus offsets for the ground truth and the tensor method are identical, and indicated in the top row, while the focus offsets of Born approximation are shifted so that the center image corresponds to the highest signal as in the case of the two rows above. From the figure, we see that the images everywhere are also in close agreement between the ground truth and the tensor approximation. On the other hand, the Born approximation shows much smaller magnitude images that also differ qualitatively from the ground truth away from the center of the image.  Figures \ref{fig:Mo200-thru-focus2}-\ref{fig:Mo200-thru-focus3} show the images computed from the scattering matrix data of larger defects in Fig. \ref{fig:Mo200} at the focus offsets indicated by dots on Fig. \ref{fig:Mo200-focus-scans}. We again see remarkable accuracy of the tensor method presented in this work in contrast to the Born approximation. 

 The right side of the table shows numerical differences in the final results, which are the images computed across various focus offsets. The errors shown there calculated as,
\begin{equation}
\epsilon(\Delta I^{approx}) = \frac{\max{|\Delta I^{approx}(\boldsymbol{r})-\Delta I(\boldsymbol{r})|}}{\max{|\Delta I(\boldsymbol{r})|}}.\label{eq:rel-err-Image}
\end{equation}
Here the maximum difference is normalized to maximum intensity in the image $\Delta I$ computed from $U_{\alpha\beta}$. The maximum is taken across the entire image and over focus offsets shown in Fig \ref{fig:Mo200-focus-scans}. It is clear that the maximum difference never exceeds a few percent in the case of tensor approximation, while it can be near or larger than 100 percent when using the Born approximation. 

I remark that while the good agreement of images is the desired end-result, the high correlation among the S-matrix elements shown in Fig. \ref{fig:Geometry} must also be emphasized. This is so because all images under any illumination or collection pattern arise from the same S-matrix. Therefore, a the correlation of the estimated S-matrix elements to the exact elements provides a very high confidence that any admissible but arbitrary imaging will produce a good agreement with exact solutions.

\begin{table}
\centering
\begin{tabular}{lrr|lrr}
\toprule
\multicolumn{3}{c|}{S-matrix Error} & \multicolumn{3}{c}{Image Error} \\
\hline
Figure & Born & Tensor& Figure & Born & Tensor \\
\midrule
\multirow{2}{*}{\ref{fig:Geometry}(a)} & \multirow{2}{*}{0.874} & \multirow{2}{*}{0.012} & \ref{fig:Mo200-focus-scans}(a) &   0.864  & 0.0084\\
                                       &                        &                        & \ref{fig:Mo200-focus-scans}(b) &   0.549  & 0.0065 \\
\hline
\multirow{2}{*}{\ref{fig:Geometry}(b)} & \multirow{2}{*}{1.035} & \multirow{2}{*}{0.060} & \ref{fig:Mo200-focus-scans}(c) &    1.030 &   0.0400 \\
                                       &                        &                        & \ref{fig:Mo200-focus-scans}(d) &    0.737 &   0.0331 \\
\hline
\multirow{2}{*}{\ref{fig:Geometry}(c)} & \multirow{2}{*}{1.066} & \multirow{2}{*}{0.086} & \ref{fig:Mo200-focus-scans}(e) &    0.914 &   0.0569 \\
                                       &                        &                        & \ref{fig:Mo200-focus-scans}(f) &    0.773 &   0.0419 \\
\bottomrule
\end{tabular}
\caption{Error in S-matrix predictions corresponding to the three cases in \ref{fig:Geometry}, and the error across images at all de-focus values shown in \ref{fig:Mo200-focus-scans}. The definitions of the errors in each column is given in equations \eqref{eq:rel-err-S} and \eqref{eq:rel-err-Image} in the text. }
\label{tab:error}
\end{table}

\newpage
\section{Discussion}\label{sec:discussion}

In the previous section, I showed how the local approximation to the susceptibility tensor leads to an accurate prediction of scattering by metallic objects of size 40 nm at the wavelength of 200 nm in free space. Note that in the language of Mie theory, the size parameter inside the defect is about $1.5$ and thus well outside the Rayleigh-Gans condition for the Born approximation to hold. In this section I expand upon the results presented above to make some general observations I have made in our numerical experiments.

Let us first discuss the overall numerical efficiency achieved in the proposed method. The efficiency arises for two reasons: (a) the reduction in the \emph{number} of numerical solutions to Maxwell's equations, and (b) the reduction in the \emph{problem size} for each solution. If we associate a unit cost $C$ to compute the solution for a single source using any method such as FDTD or other approaches \cite{Vettenburg:19}, then the cost to compute $n_d$ columns of the S-matrix is essentially $n_d C$. The efficiency gain in (a) arises from replacing $n_d$ by a much smaller number. 

Similarly, the unit cost $C$ increases in proportion to the area $A$ of the computational domain perpendicular to the surface normal. Any direct calculation of S-matrix requires $A$ to increase  inversely to the resolution in the space of scattering directions. The efficiency gain in (b) arises by reducing $C$ by replacing $A$ with a much smaller area that is large enough to minimize inter-defect scattering due to periodic boundary conditions. Therefore (b) is of most interest in absorptive materials where extinction within the material effectively introduces a decay of inter-defect coupling with distance. When extinction is not sufficient a large domain must be simulated and the number of outgoing directions increases for a single source. In that case reciprocity theorem implies that  (a) can essentially reduce $n_d$ to 1 since larger area increases the number of outgoing directions resolving the scattered light. 

Therefore, the efficiency gain is independent of how fast the underlying near field solver works, and depends only on how many times it must be run. A faster solver such as \cite{Vettenburg:19} would simply reduce $C$, but (a) will reduce the product $n_d C$ as discussed above, and in more detail below.

The method becomes advantageous in proportion to the number of sources, or illumination directions, that must be simulated in order to achieve the final results. This is so because the background fields can be computed quickly in the smallest unit cell, or the Wigner-Seitz cell of the periodic structure, and over a dense sampling, $n_d$, of the pupil. To generate the training and validation data, it suffices to have about 100 scattering matrix elements. This number is based on my experience in applying this method to a diverse set of structures ranging from the back-end of line metal layers as shown here, to defects in nano-sheet transistors \cite{nanosheet} in a static random access memory layout, metallic connections in storage cells of dynamic access memory, and channels in three-dimensional NAND flash memory.

The training cell can be a large supercell where only one illumination direction can produce a 100 diffraction orders, or it can be a smaller cell as was the case here, where 10-20 illumination directions are simulated to produce a desired dataset. In the former case, the total reduction in the computational effort is a factor $n_d$ since only one calculation enables the tensor approximation over the entire illumination space with the same supercell. There is extra cost of storing the near fields inside the defect volumes and computing the overlap integrals. By storing instead the projections of these integrals to the basis functions, I have found this to incur little extra computational time. 

In the second case, let $n_t$ be the number of ab-initio computations done with a smaller supercell of area $A_t$ as compared to the larger area $A_s$ needed to directly obtain $n_d$ diffraction orders per illumination direction. To have $n_d$ samples in a k-space disk of radius $2\pi\lambda$, we must then have $A_s > n_d \lambda^2$. If $C$ is the cost to compute one column of the S-matrix with area $A_s$, then the total cost for a fully ab-initio computation is $C_s = n_d C$. With the tensor approximation, the we would expect the cost to scale at least linearly with the supercell area on a shared memory machine\footnote{the reduction would be much higher if $A_s$ requires a distributed memory machine while $A_t$ is able to fit in a single node of a distributed computing cluster.}. Thus the cost with the tensor approximation is expected to be $C_t < n_t A_t C_s / A_s$, and substituting the above estimates, 
\begin{equation}
    C_t < \frac{n_t A_t}{n_d^2 \lambda^2}C_s.
\end{equation}

In structures with absorptive materials surrounding the defect, $A_t < \lambda^2$ suffices when the effective absorption length scale is below one wavelength. In that case the cost scales as $n_t / n_d^2$, which is enormous. The denominator of $n_d^2$ follows because one factor is arising from reduction in the number of independent sources simulated, while another factor arises from the reduction in the domain area simulated. Thus with $n_t \approx 20$ and $n_d \approx 100$, we can reduce the computation by a factor of 500. This reduction in the time or cost can be used to explore the parameter space with variations of the domain geometry and materials. We would thus expect significant value of this approach to the applications of this to search based inverse design problems.

The choice of training data requires further comments. The predicted S-matrices tend to inherit the symmetry properties of the training data. Thus for domains that possess a point-group symmetry, a good choice is to generate the training data from the set of incidence wavevectors ${\boldsymbol{k}'}$ that is closed under this group. Since this symmetry is rotational symmetry around the surface normal, this essentially results in a choice of a few wavevectors in the \emph{irreducible} Brillouin zone of the periodic lattice of the background structure, and then applying all elements of the symmetry group to generate the full training data. With typical patterns having 2- to 6-fold rotational symmetry, this choice of training data lead to  $n_t \approx 20$. Furthermore, the chosen points still respect the translational symmetry of the supercell $A_t$ in order for the outgoing wavevectors to share the in-plane components with the incident set $\{\boldsymbol{k}'\}$. This allows an immediate application of reciprocity theorem without generation of an extra set of fields with time-reversed directions $\boldsymbol{k}\rightarrow-\boldsymbol{k}$ . In the future, it is worth also exploring the use of other types of sources, such as an array of dipole sources, to generate $U_{\alpha\beta}$
 and $\tilde{U}_{\alpha\beta}$ for estimating the tensor. 
 
Furthermore, only the local approximation is used in this work. The approximation suffices in all cases of interest studied here, with the size parameters in the range of 1-2. However, detection of larger objects in structured media is still important. Addressing these larger objects will require non-local terms to be estimated, and a careful construction of the basis functions to minimize the computational effort.

Finally, the question arises as to what the spatial dependence of $\Lambda_{ij}(\boldsymbol{r})$ is and whether it is sufficient to generate the correct near field in the defect volume. Generally, the integral quantities presented above (scattering matrices and energy) are predicted accurately with a low order of the basis functions, but near fields require this basis to be enlarged. Since the objective in many applications is these integral quantities, and those are the ones studied in this work, the efficiency of this method is proved only for these types of calculations. The use of this technique to faithfully construct near fields inside the defect at significantly lower cost is yet to be explored.

\section{Conclusion}

I have presented a numerical parameter estimation approach to the extension of the Born approximation for periodic dielectric media. This approach is formally based on the Born series expansion of the electric field integral equation with a finite sized dielectric perturbation imposed on the media. I demonstrated through direct comparison of the scattering matrices computed ab-initio that our approach retains the accuracy while reducing the computational effort by at least two orders of magnitude. The scattering matrix elements predicted by this approach retain high accuracy in both the magnitude and the phase, and therefore predict the correct through focus images from the scattered fields. I computed the latter through a direct application of the Abbe theory to the scattering matrices.

This work shows that when considering a restricted set of quantities, statistical estimation can be used to effectively replace a large number of costly solutions to Maxwell equations in complex dielectric and metallic media. I believe that with the method proposed in this work, this approach to extending the Born approximation can be applied in many inverse design problems. The speedup gained, even with less accuracy than presented here, can increase the search capacity by two orders of magnitude simply due to the reduction in the necessary ab-initio solutions to the Maxwell equations. 

\section*{Acknowledgments}
I acknowledge many in-depth discussions with Apo Sezginer, Stanley Burgos, and Vaibhav Gaind. I also thank Apo Sezginer for bringing to my attention the work by Habashy and co-workers during the course of this work. 

\section*{Disclosures} Kuljit S. Virk: KLA Corporation (I,E).

\appendix
\section{Source terms}\label{app:sources}
Here I give an expression for the polarization sheet at $z=z_s$ that radiates a plane wave into the half space below it. If the amplitude of the electric field is $E_0$ and its electric polarization vector is $\hat{e}(\boldsymbol{k})$, then 
\begin{equation}
    \boldsymbol{P}^s(\boldsymbol{r},\omega)  = \frac{2\epsilon_{0}k_{z}}{ik^{2}}E_0
    e^{i\boldsymbol{\kappa}\cdot\boldsymbol{r}_\perp}\delta(z-z_s)\hat{e}(\boldsymbol{k}).
\end{equation}
Here $\boldsymbol{\kappa}$ is the in-plane component of $\boldsymbol{k}$ and $k_z=\sqrt{k^2-\kappa^2}$. By applying the free space Green function to this expression, it can be easily verified that this polarization generates the field $E_0 \hat{e}(\boldsymbol{k})e^{i(\boldsymbol{\kappa}-k_z\hat{z})\cdot\boldsymbol{r}}$. The most convenient form of the Green function, which is Fourier transformed in the plane of the sheet can be found in \cite{sipe1987}.

\section{Integral equation for the electric field}\label{app:E-equation}
The total field satisfies Eq. \eqref{eq:total-efield-eq} The operator acting on the total Green function on the left hand side is the inverse zeroth-order Green function, and thus the equation may be re-written as,
\begin{equation}
    E_{i}(\boldsymbol{r}) = 
    E^{(0)}_i(\boldsymbol{r})+
    k^2 \int \mathrm{d}^3\boldr_1 G^{(0)}(\boldsymbol{r},\boldsymbol{r}_1)\chi^{(d)}_{il}(\boldr_1)E_{l}(\boldr_1).\label{eq:IE-G-appendix}
\end{equation}
In order to simplify the expressions below, it is convenient to introduce infinite dimensional vector $\underline{E}$, and matrices $\widehat{G}$ with matrix elements $G_{ij}(\boldsymbol{r},\boldsymbol{r}')$, and a matrix $\widehat{\chi}^{(d)}$ with matrix elements $\chi_{ij}(\boldsymbol{r})\delta(\boldsymbol{r}-\boldsymbol{r}')$. In terms of these matrices, \eqref{eq:IE-G-appendix} becomes,
\begin{equation}
    \underline{E} = \underline{E}^{(0)} + k^2\widehat{G}^{(0)}\widehat{\chi}^{(d)}\underline{E}\label{eq:IE-G-matrix-appendix}
\end{equation}
I substitute the formal solution to this equation on the right hand side,
\begin{equation}
    \underline{E} = \widehat{E}^{(0)} + 
    \widehat{G}^{(0)}\widehat{\chi}^{(d)}
    \left[I-k^2\widehat{G}^{(0)}\widehat{\chi}^{(d)}\right]^{-1}
    \underline{E}^{(0)}\label{eq:IE-G-matrix-appendix2}
\end{equation}
Based on this equation, I introduce the operator $\mathcal{\widehat{R}}^{(d)}$ that satisfies the equation,
\begin{equation}
\mathcal{\widehat{R}}^{(d)} = I + k^2\widehat{G}^{(0)}\widehat{\chi}^{(d)}\mathcal{\widehat{R}}^{(d)},
\end{equation}
and thus generates the total field in the presence of a defect from the background field. Restoring the matrix element form gives the expression in \eqref{eq:resolvent-IE-1}.
\section{Statistical estimation terminology}\label{sec:terminology}

This work combines deterministic computational approach of electromagnetism with statistical estimation. The latter enters in the strategy described above where numerical solutions to Maxwell's equations are used to estimate the parameters to compute a scattering matrix. I discuss this estimation by borrowing terminology from statistical learning, specifically, the concepts of training set, validation set, ground truth and (over-) fitting. To serve as a self-contained reference for readers who may be less familiar with these terms, I define them precisely what they mean in the context of this paper.  

To be precise, I will define a tuple $t_{\alpha}=(\boldsymbol{P}^s_\alpha,\boldsymbol{E}^{(0)}_\alpha,\boldsymbol{E}_\alpha)$, which contains the source and the field solutions in the absence and the presence of a defect. The term \emph{training set} means a set of $t_\alpha$ used to compute $U_{\alpha\beta}$ in \eqref{eq:Uexact} directly from the field solutions. This matrix enters the left hand side of \eqref{eq:U-tilde}, and the tensors on the right hand side of this equation are computed from the corresponding $\boldsymbol{E}^{(0)}_\alpha$ as specified by Eqs. \eqref{eq:E-phi-voronoi}-\eqref{eq:W-voronoi}. This data is substituted into  \eqref{eq:U-tilde} to solve for an \emph{estimate} of the tensor elements $\Lambda_{ij}$. The latter are also referred to as \emph{parameters} of the model of the S-matrix, \textit{i.e.} Eq. \eqref{eq:U-tilde}. This estimation procedure is typically referred to as \emph{training} in the practice of statistical learning.
 
The term \emph{validation set} refers to the set of $t_\alpha$ for which all the above data is computed \emph{but} not used when estimating $\Lambda_{ij}$ in \eqref{eq:U-tilde}. Instead the estimated parameters $\Lambda_{ij}$ are substituted into  the right hand side of \eqref{eq:U-tilde}, and the $\alpha$-dependent quantities are taken from the validation set. This generates a prediction for $U_{\alpha\beta}$ by the model equation \eqref{eq:U-tilde}. Since $U_{\alpha\beta}$  are also computed directly for the validation set as well, the error in this prediction can be computed. Keeping this so-called \emph{validation error} low reduces the over-specialization or \emph{over-fitting} the parameters $\Lambda_{ij}$ to the training set.

Another statistical term used below is the \emph{ground truth}, which refers to $U_{\alpha\beta}$ computed directly but not used in training. Instead these quantities are used to test the \emph{generalization} of $\Lambda_{ij}$ from the training set to any other set of sources. The term ground truth is also used for images computed from these "ground truth" scattering matrices. 

Finally, the predicted results are compared to the ground truth by plotting them against each other in the case of S-matrices. A quantitative measure of how well the prediction matches the ground truth is the correlation coefficient $R$ for regression of the prediction against the ground truth \cite{Bulmer}.  

\section{Abbe Image Calculation}\label{sec:ImageCalc}

In order to interpret the relevant accuracy of the fitting procedure for susceptibility, I computed incoherent images from the resulting S-matrix elements. The intensity in the image plane at distance $\delta z$ is computed from the Abbe theory \cite{kitkwong} using the formula 
\begin{equation}
I[S](\boldsymbol{x}) = \int d^2\boldsymbol{\kappa}'\left|\int d^2\boldsymbol{\kappa}%
   \left({\frac{1-\kappa^2/k^2}{1-\kappa^2/M^2/k^2}}\right)^{1/4} 
   e^{i\boldsymbol{\kappa}\cdot\boldsymbol{x}+ik_z \delta z}
    \hat{e}_{\alpha}(\boldsymbol{\kappa}) S_{\alpha\beta}(\boldsymbol{\kappa},\boldsymbol{\kappa}') \hat{e}_\beta(\boldsymbol{\kappa}')\cdot\boldsymbol{E}(\boldsymbol{\kappa}') \right|^2.   \label{eq:abbe}
\end{equation}
In this equation $k_z = \sqrt{k^2 - \kappa^2/M^2}$, where $\boldsymbol{\kappa}$ is the projection of the wavevector $\boldsymbol{k}$ onto the surface of the structure, and $M$ is the magnification factor set to 100 in our calculations. The function $\boldsymbol{E}(\boldsymbol{\kappa}')$ are the vector amplitudes of the incident plane waves. The collection of plane waves is open within $|\boldsymbol{\kappa}|\leq 0.9 k$.  

 The discussion in the text refers to the difference $\Delta I = I[S_0+\Delta S]-I[S_0]$, where $S_0$ is the S-matrix of the background and $\Delta S$ is the perturbation caused by the defect. This so-called difference image carries the information from the field scattered by the defect. Under the so-called bright-field imaging, this field interferes with the background, while under dark-field imaging, collection pupil is modified to collect only the intensity from the scattered field alone. This closely represents the methodology of detecting sub-wavelength objects as intensity anomalies in the difference of images.

\bibliography{BibCollection}

\end{document}


\maketitle

\section{Ab-initio Calculation of Near Fields}

The internal electric fields discussed in the main text are calculated using the Finite Difference Time Domain (FDTD) method implemented in the package MEEP \cite{meep}. Bloch periodicity is employed in the plane of the structure, the x-y plane, and perfectly matched layers (PMLs) at the top and the bottom of the structure. Excitation was created with a current sheet  $\boldsymbol{\hat{j}}e^{i\boldsymbol{\kappa}\cdot\boldsymbol{r}_\perp-i\omega t}p(t)$, where $\boldsymbol{\kappa}$ is the projection of the wavevectors onto the plane of the structure. The source is multiplied by a finite Gaussian time-dependent pulse, $p(t)$, with center frequency corresponding to 200 nm and bandwidth to 50 nm. The calculations were run until the magnitude of the field in the upper-half-space vanished below  $10^{-4}$. The resulting time-dependent fields were analyzed using harmonic inversion \cite{meep} and the field components for the frequency corresponding to the wavelength of 200 nm were extracted.

To account for the strong dispersion of the materials in the frequency regime of interest, MEEP couples the Maxwell equations to polarization source terms that are themselves driven by the electric fields \cite{meep}. The total polarization is modeled as a sum of simpler polarization fields each evolving as a driven harmonic oscillator. In frequency domain, this corresponds exactly to a linear dielectric function with the Lorentzian dispersion, 

\begin{equation}
    \varepsilon(\omega) = \left(1 + \frac{i\sigma_D}{2\pi\omega}\right )
    \left[\varepsilon_\infty + \sum_{n=1}^{N} \frac{\omega_n^2 \sigma_n} {\omega_n^2 - \omega^2 -i \omega \gamma_n}\right]\label{eq:dielectricmodel}
\end{equation}

\begin{figure}[bh]
\centering
\begin{subfigure}[b]{0.4\textwidth}
\includegraphics[height=1.6in]{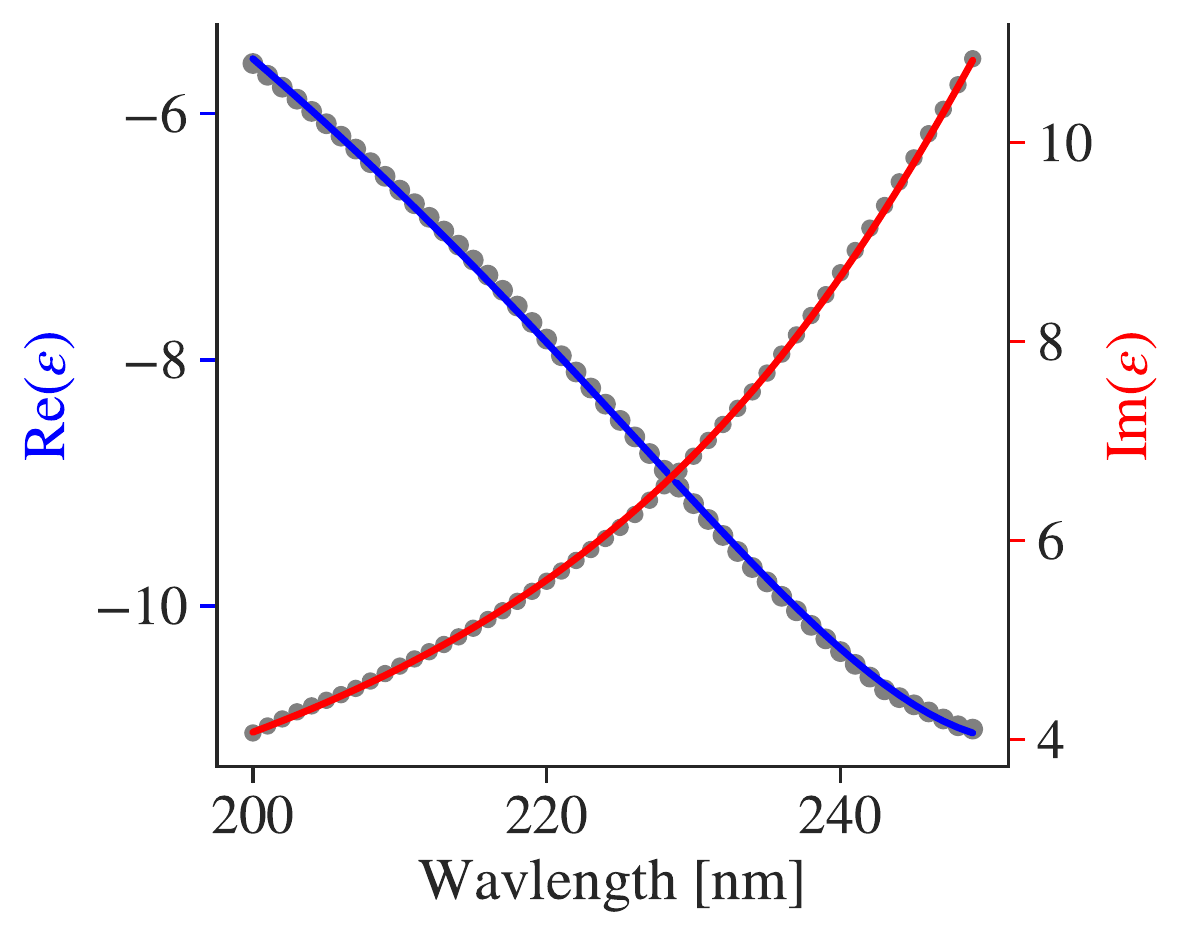}
\subcaption{}
\end{subfigure}
\begin{subfigure}[b]{0.4\textwidth}
\centering
\includegraphics[height=1.6in]{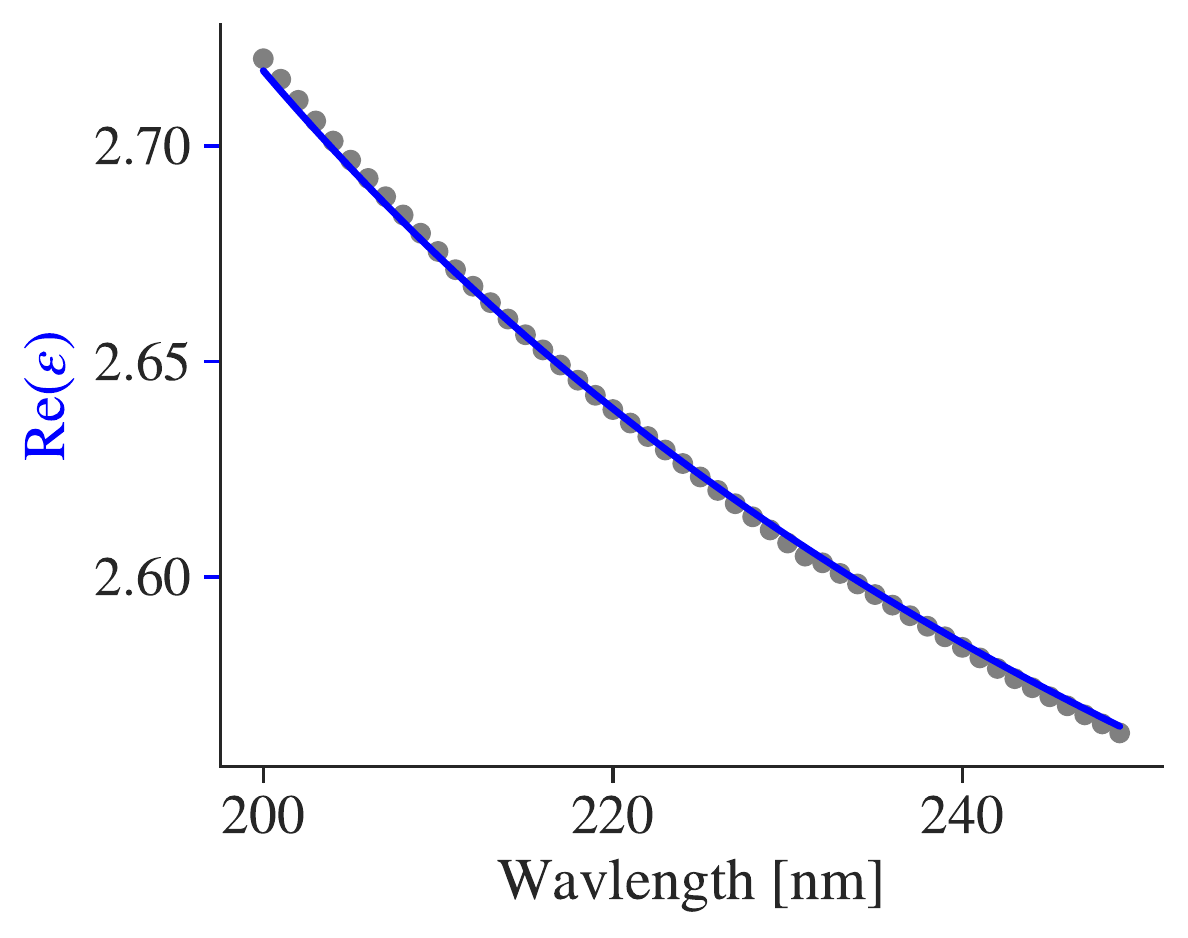}
\subcaption{}
\end{subfigure}

\caption{(a) Real and imaginary parts of the dielectric function of Mo. The gray dots are data from Palik while the solid lines are produced by the analytical model \eqref{eq:dielectricmodel} with $N=5$. (b) Real part of the dielectric function of SiO2 with data from Palik (dots) and the Lorentzian model fit with $N=1$.}

\label{fig:oscillators}
\end{figure}

The parameters of this oscillator model are fit to the optical constants data \cite{palik} over the limited range of wavelengths from 195 nm to 300 nm. I performed model selection by varying $N$ from 1 to 5 to obtain the best fit results. The results for Mo and SiO$_2$ are shown in figure \ref{fig:oscillators}. Mo data corresponds to $N=5$ and SiO$_2$ data to $N=1$. I show the parameters for the metal lines, Mo, in Table \ref{tab:oscillator parameters}.

As discussed in the text, the ground truth scattering matrix of the defect is generated using the exact overlap integral between the time-reversed background field $\boldsymbol{E}^{(0)}$ and the induced polarization inside the defect volume. Due to the periodic boundary conditions in the lateral dimensions, these dimensions must be large enough to suppress the effect of neighboring defects placed periodically. Letting this dimension be $L$, we write the scattering matrix as,
\begin{eqnarray}
   \Delta S_L(\boldsymbol{k}_\alpha,\hat{e}_\alpha;\boldsymbol{k}'_\beta,\hat{e}_\beta) = \frac{-ik^2}{2 E_0 \epsilon_0 k_z} \sum_i \int_{V^d} \mathrm{d}^3\boldsymbol{r} \boldsymbol{E}_L^{(0)}(\boldsymbol{r};-\boldsymbol{\kappa},\hat{e}) \boldsymbol{P}_L^{(d)}(\boldsymbol{r};\boldsymbol{\kappa}',\hat{e}').\label{eq:smatrix}.
\end{eqnarray}
The exact scattering matrix of the periodic array of defects is 
\begin{equation}
    \Delta S_{per}(\boldsymbol{k}_\alpha,\hat{e}_\alpha;\boldsymbol{k}'_\beta,\hat{e}_\beta) = \Delta S_L(\boldsymbol{k}_\alpha,\hat{e}_\alpha;\boldsymbol{k}'_\beta,\hat{e}_\beta) \sum_{\boldsymbol{R}} e^{i(\boldsymbol{\kappa}-\boldsymbol{\kappa}')\cdot \boldsymbol{R}},
\end{equation}
where $\boldsymbol{R}$ are the lattice vectors and the summation represents the structure factor in the form of delta functions located in the wave-vector space at the reciprocal lattice vectors. However, if the structure factor is ignored, then the scattering matrix of an \emph{isolated defect} is obtained in the limit, $L\rightarrow\infty$, \emph{i.e..},

\begin{equation}
    \Delta S_{iso}(\boldsymbol{k}_\alpha,\hat{e}_\alpha;\boldsymbol{k}'_\beta,\hat{e}_\beta) = \lim_{L\rightarrow\infty} \Delta S_L(\boldsymbol{k}_\alpha,\hat{e}_\alpha;\boldsymbol{k}'_\beta,\hat{e}_\beta)\end{equation}

Note that due to the overlap integral in the definition of $\Delta S_L$ being restricted to lie inside the defect volume, the field solutions $\boldsymbol{E}_L$ must converge only inside the defect volume, placed farthest from the boundaries, and it must converge only in the integral sense and not necessarily point-wise. In structures with sufficiently high absorption, $L$ need not be very large before convergence is achieved. Furthermore, the approach to this limit does not impact the ability to fit the parameters of the susceptibility tensor. It only leads to an overall scaling. In this work, I found that $L > 100 nm$ suffices for the Mo/SiO$_2$ line space system at a wavelength of 200 nm. When the length scale for the scattered field inside the structure is governed by absorption, the wavelength plays only a secondary role in determining the dimensions of the domain through the dispersion relation $\epsilon(\omega)$. This is another direct benefit of using the formula \eqref{eq:smatrix} to compute the scattering matrix in contrast to computing it by the difference of scattering matrices in the presence and absence of the defect. The latter require the domain size to be directly governed by the number of propagating modes allowed by the Bloch periodicity of the structure.

\begin{table}
\centering
\begin{tabular}{lrrr}
\toprule
{} &  $\omega_n/2\pi$ &  $\gamma_n/2\pi$ &  $\sigma_n$ \\
\midrule
0 &    0.001964 &    0.000198 &   27.496889 \\
1 &    0.003658 &    0.000912 &    3.057252 \\
2 &    0.006121 &    0.013024 &    1.531705 \\
3 &    0.008171 &    0.059215 &    7.372552 \\
4 &    0.010273 &    0.086407 &    0.690262 \\
\bottomrule
\end{tabular}
\caption{Parameters of the Lorentz oscillator model \eqref{eq:dielectricmodel} for Mo. The fit results are shown in \ref{fig:oscillators}. The parameter $\varepsilon_\infty = 2.08$ and $\sigma_D=0$.}\label{tab:oscillator parameters}
\end{table}
